\documentclass{article}

\usepackage{amsfonts}
\usepackage{ulem}
\usepackage{float}
\usepackage{amssymb}
\usepackage{amsmath}
\usepackage{mathrsfs} 
\usepackage[all]{xy}
\usepackage[dvips]{color}
\usepackage[dvips]{graphicx}

\usepackage{hyperref}

\newcommand{\omits}[1]{}

\definecolor{dyellow}{rgb}{1.,0.8,.0}
\definecolor{myblue}{rgb}{.1,.1,.7}
\definecolor{dcyan}{rgb}{.0,.6,.6}
\definecolor{dmagenta}{rgb}{0.6,0.0,0.6}
\definecolor{brown}{rgb}{0.6,0.2,0.}
\definecolor{darkblue}{rgb}{.0,.0,0.5}
\definecolor{darkred}{rgb}{0.75,0.0,0.0}
\definecolor{orange}{rgb}{1.,.6,.0}
\definecolor{dorange}{rgb}{0.8,.4,.0}
\definecolor{darkgreen}{rgb}{0.0,0.6,0.0}
\definecolor{purple}{rgb}{.4,.0,.4}
\definecolor{grey}{rgb}{0.7,0.7,0.7}

\newcommand{\hide}[1]{}
\newcommand{\delete}[1]{}
\newcommand{\distribute}[3]{}	

\newcommand{\od}{\mathrm{d}}

\title{\textbf{Hamiltonian Analysis of 3-dimensional Spacetime in Bondi-like Coordinates}}

\author{Chao-Guang Huang$^{1}$\thanks{Email: \text{huangcg@ihep.ac.cn}}\quad
and\quad Shi-Bei Kong$^{1,2}$\thanks{Email: \text{shibeikong@ecut.edu.cn}}
\\
\\
$^{1}$Theoretical Physics Division, Institute of High Energy Physics, CAS, Beijing, China 100049
\\
$^{2}$University of Chinese Academy of Sciences, Beijing, China 100049}

\begin{document}

\maketitle

\begin{abstract}

The Hamiltonian analysis for a 3-dimensional connection dynamics of $\frak{so}(1,2)$, spanned by $\{L_{-+},L_{-2},L_{+2}\}$ instead of $\{L_{01}, L_{02}, L_{12}\}$, is first conducted in a Bondi-like coordinate system. The symmetry of the system is clearly presented.
A null coframe with 3 independent variables and 9 connection coefficients are treated as basic configuration variables. All constraints and their consistency conditions, the solutions of Lagrange multipliers as well as the equations of motion are presented.
There is no physical degree of freedom in the system. The Ba\~nados-Teitelboim-Zanelli (BTZ) spacetime is discussed as an example
to check the analysis. Unlike the ADM formalism, where only non-degenerate geometries on slices are dealt with and the Ashtekar formalism,
where non-degenerate geometries on slices are mainly concerned though the degenerate geometries may be studied as well,
in the present formalism the geometries on the slices are always degenerate though the geometries for the spacetime are not degenerate.

\end{abstract}

\tableofcontents

\section{Introduction}

How to establish a theory of quantum gravity is a long-standing problem in theoretical physics.
Among many conceptual and technical problems, the choice of basic configuration variables is an important one.
The ADM formalism \cite{ADM} provides a formulation to study the initial-value problem in general relativity
and to quantize the Einstein theory of gravity. The formalism foliates a 4-dimensional spacetime into a series of 3-dimensional
spacelike hypersurfaces along an arbitrary timelike direction. Each 3-dimensional spacelike hypersurface is labelled by a timelike coordinate.
The basic configuration variables in the formalism are the components of the 3 dimensional induced metric on any spacelike hypersurface.
The canonical quantization leads to the Wheeler-De Witt (WDW) equation \cite{WD}, which is the basic equation for the first non-perturbative theory
of quantum gravity but too complex to get a non-trivial exact solution.

After the Ashtekar variables are introduced \cite{Ashtekar}, general relativity can be reformulated as an $\mathfrak{su}(2)$-connection dynamics
(or ${\frak so}$(3)-connection dynamics). In the Ashtekar's formalism, the 4 dimensional spacetime is also foliated into a series
of 3 dimensional spacelike hypersurfaces along a timelike direction. There is a local SO(3)-rotation symmetry at any point
in an arbitrary 3-dimensional spacelike hypersurface. The self-dual $\mathfrak{su}$(2)-connection (or self-dual $\frak{so}$(3)-connection)
are chosen as the basic configuration variables, and the densitized 3 frame fields as their conjugate momenta.
In the Ashtekar's formalism the (quantum) constraint equations are transformed into the form of polynomials, which can be easily solved.
Based on the Ashtekar variables, the loop quantum gravity has been established.

The Ashtekar formalism, however, only applies to the study of 4-dimensional gravitational theories  because in a 4-dimensional spacetime the local
(or internal) symmetry SO(1,3) can be decomposed into the direct product of two SO(3),
but in a higher dimensional spacetime the similar decomposition does not exist.
In order to generalize the Ashtekar formalism to a higher-dimensional spacetime, Bodendorfer, Thiemann, and Thurn (BTT) suggest to choose
${\frak so}(d)$-connection instead of ${\frak so}(d-1)$-connection as the basic configuration variables in a $d$-dimensional (Lorentz) spacetime \cite{BTT1,BTT2,BTT3,BTT4,BTT5}, which is a highly non-trivial method. Although the Hamiltonian formalism based on the configuration variables 
can be established in a $d$-dimensional Lorentz spacetime \cite{BTT1}, the Lagrangian formalism on the spacetime fails to be constructed \cite{BTT2}.
Only when the spacetime is an Euclidean one, both the Lagrangian and Hamiltonian formalisms are valid at the same time.

Recently, in the study of the statistical origin of black hole entropy, it has been shown from the Lagrangian formalism
that an ${\frak so}(1,1)$ BF theory can always be acquired as the limit of ${\frak so}(1,d-1)$-connection on an isolated horizon 
\cite{BF1,BF2,BF3,BF4,BF5,BF6} when $B$ field (a $d-3$-form field) is defined by
${\rm d} B=\Sigma_{-+}=\Sigma_{01}=1/(d-2)!\epsilon_{01K...N}e^K\wedge...\wedge e^N$ and SO(1,1) gauge field $F$ is defined
by $F=F^{-+}$, the $-+$ part of the SO(1,$d-1$) gauge field $F^{IJ}$, on the horizon.
One of the starting points to obtain  a boundary ${\frak so}(1,1)$-BF theory is to choose the Bondi-like coordinate system near an isolated horizon.
In the Bondi-like coordinate system, a lightlike coordinate $v$ is chosen as the ``time" coordinate, instead of a timelike coordinate $t$ as usual
in the $1+(d-1)$-decomposition. In other words, the spacetime is foliated into a series of $(d-1)$-dimensional null
hypersurfaces along a lightlike evolution direction. In the explanation of the statistical origin of the entropy of an isolated horizon,
on the other hand, the quantum states are still calculated from the loop quantum gravity based on the Hamiltonian formalism constructed by BTT.
To make the explanation more self-consistent, one needs to re-analyze the bulk quantum states based on a formalism which can approach the boundary
${\frak so}(1,1)$-BF theory. The purpose of the present paper is to make the first step to construct such a theory.
In this paper, the canonical formalism in a 3-dimensional spacetime, using a lightlike or null coordinate as the evolution coordinate, is established.
As a result, the geometries on the slices are always degenerate though the geometries for the spacetime are not degenerate.
This is very different from the ADM formalism in which only non-degenerate geometries on slices are dealt with [1], [2]
and from the Ashtekar formalism in which non-degenerate geometries on slices are mainly concerned though the degenerate geometries may be studied as well
\cite{Rovelli,Ma-Liang-Kuang,Ma-Liang}.

In the literature, there have been several efforts to make a 3+1 decomposition of a 4-dimensional spacetime along a lightlike direction \cite{GRS}, \cite{Goldberg} or a 2+2 decomposition along double null directions \cite{d'I-L-V,d'I-L-V2,AS}.
However, in these efforts the ${\frak so}(3)$ connection dynamics is still built. By these approaches,
${\frak so}(1,1)$-connection cannot be obtained obviously on an isolated horizon.
Our key observation is that a Lorentz algebra $\mathfrak{so}(1,d-1)$ can be decomposed into $\mathfrak{so}(1,1)\oplus\mathfrak{so}(d-2)\oplus\mathfrak{t}^{-}(d-2)\oplus\mathfrak{t}^{+}(d-2)$,
where $\textrm{t}^{\pm}(d-2)$ are the translation algebras in $(d-2)$-dimensional spaces\cite{Hall,BCG}.
$\mathfrak{so}(1,1)$, $\mathfrak{so}(d-2)$, $\mathfrak{t}^{-}(d-2)$, and $\mathfrak{t}^{+}(d-2)$
are all subalgebras of $\mathfrak{so}(1,d-1)$, and correspondingly, Lie groups SO$(1,1)$, SO$(d-2)$, $\textrm{T}^{-}(d-2)$,
and $\textrm{T}^{+}(d-2)$ are all subgroups of the Lie group SO$(1,d-1)$.
If a connection dynamics is based on this kind of decomposition, the boundary SO$(1,1)$-BF theory can be acquired naturally in an arbitrary
dimensional spacetime. The other purpose of the present paper is to investigate the feasibility of the connection dynamics
based on this kind of decomposition. As a simple example, the 3-dimensional Palatini action is considered.

The arrangement of the paper is as follows. In Sec.2, the local symmetry, metric and coframe in the Bondi-like coordinate system, the Palatini action and Hamiltonian of a 3-dimensional spacetime, and the primary constraints are briefly introduced.
In Sec.3, the consistency conditions for the primary and the secondary  constraints as well as the degree of freedom are analyzed.
In Sec.4, all equations of motion are presented. In Sec.5, Ba\~nados-Teitelboim-Zanelli (BTZ) spacetime \cite{BTZ} is used to examine the analysis.
In the last section, some concluding remarks are given.

In this paper, the natural unit system is used, where $c=8\pi G=1$.

\section{Preliminary}

\subsection{Local Symmetry}

A 3-dimensional spacetime has a local or internal Lorentz symmetry, described by the Lie group SO(1,2).
The local SO(1,2) transformations leave the metric of the 3-dimensional spacetime invariant.
The SO(1,2) transformations are usually classified into 3 basic transformations, namely 2 boosts and 1 rotation.
The generators of SO$(1,2)$ can be realized by the following form,
\begin{alignat}{1} \label{am}
L_{IJ}=x_{I}\frac{\partial}{\partial x^{J}}-x_{J}\frac{\partial}{\partial x^{I}}\omits{=x_{I}\partial_{J}-x_{J}\partial_{I}},\qquad x_{I}=\eta_{IJ}x^{J},
\end{alignat}
where $\{x^{I},\ I=0,1,2\}$ are local Minkowski coordinates in 3 dimensional spacetime and $\eta_{IJ}=\textrm{diag}(-1,1,1)$.
There are 3 independent generators, $L_{01},L_{02}, L_{12}$, the former 2 generate boost transformations, and the last one generates a rotation.
The Lie brackets of the generators are
\begin{alignat}{1}
[L_{IJ},L_{KL}]=\eta_{IL}L_{JK}+\eta_{JK}L_{IL}-\eta_{IK}L_{JL}-\eta_{JL}L_{IK}.
\end{alignat}
With the help of $L_{IJ}$, 3 new generators for the $\frak{so}(1,2)$ algebra can be defined\cite{Hall},
\begin{alignat}{1}
&L_{-+}:=L_{01},\qquad L_{-2}:=\frac{1}{\sqrt{2}}(L_{02}-L_{12}),\qquad
L_{+2}:=\frac{1}{\sqrt{2}}(L_{02}+L_{12}).
\end{alignat}
The Lie brackets of the new generators are
\begin{alignat}{1}
[L_{-+},L_{-2}]=-L_{-2},\quad [L_{-+},L_{+2}]=L_{+2},\quad [L_{-2},L_{+2}]=-L_{-+}.
\end{alignat}
Thus, $L_{-2}$ and $L_{+2}$ may be regarded as the generators of two 1-dimensional translation algebras $\frak{t}^{-}(1)$ and $\frak{t}^{+}(1)$, respectively.
Now the algebra $\mathfrak{so}(1,2)$ is spanned by $\{L_{-+},\ L_{-2},\ L_{+2}\}$.
In this decomposition, internal indices $I,\ J,\ \cdots$ are labelled by $\{-,\ +,\ 2\}$ instead of $\{0,\ 1,\ 2\}$.

This kind of decomposition can be easily generalized to a higher dimensional spacetime.
A $d$-dimensional spacetime has a local $\textrm{SO}(1,d-1)$ symmetry, the generators of the algebra $\frak{so}(1,d-1)$ can be defined by
\begin{alignat}{1}
&L_{-+}:=L_{01},\quad \
L_{-A}:=\frac{1}{\sqrt{2}}(L_{0A}-L_{1A}),\quad \
L_{+A}:=\frac{1}{\sqrt{2}}(L_{0A}+L_{1A}),\quad \
L_{AB}:=L_{AB},
\end{alignat}
where $A,\ B=2,\ ...,\ d-1$ and $L_{AB}$ take the same form as \eqref{am}.
Their Lie brackets are
\begin{alignat}{1}
[L_{-+},L_{-A}]=&-L_{-A}, \quad [L_{-+},L_{+A}]=L_{+A}, \quad [L_{-+},L_{AB}]=0, 
\nonumber \\
[L_{-A},L_{-B}]=&0, \quad [L_{-A},L_{+B}]=L_{AB}-\eta_{AB}L_{-+}, \quad [L_{-A},L_{BC}]=\eta_{AB}L_{-C}-\eta_{AC}L_{-B}, 
\nonumber \\
[L_{+A},L_{+B}]=&0, \quad [L_{+A},L_{BC}]=\eta_{AB}L_{+C}-\eta_{AC}L_{+B}, 
\nonumber \\
[L_{AB},L_{CD}]=&\eta_{AD}L_{BC}+\eta_{BC}L_{AD}-\eta_{AC}L_{BD}-\eta_{BD}L_{AC}.
\end{alignat}
The commutation relations show that the algebra $\mathfrak{so}(1,d-1)$ may be decomposed
as
\begin{alignat}{1}
\mathfrak{so}(1,d-1)=\mathfrak{so}(1,1)\oplus\mathfrak{so}(d-2)\oplus\mathfrak{t}^{-}(d-2)\oplus\mathfrak{t}^{+}(d-2).
\end{alignat}
In particular, in a 4-dimensional spacetime,
\begin{alignat}{1}
\mathfrak{so}(1,3)=&\mathfrak{so}(1,1)\oplus\mathfrak{so}(2)\oplus\mathfrak{t}^{-}(2)\oplus\mathfrak{t}^{+}(2).
\end{alignat}
In the following part of the paper, only the most simple case is discussed, where $d=3$.

\subsection{Metric and Coframe}

For a 3-dimensional spacetime, the most general form of the metric is
\begin{equation}
\od s^2 = g_{00}(\od x^0)^2 + 2g_{0i} \od x^0 \od x^i +g_{ij}\od x^i \od x^j,
\end{equation}
where $i,j =1, 2$.  There are 6 independent components.  The 3-dimensional vacuum Einstein field equations,
\begin{equation}
R_{\mu \nu} -\frac 1 2 R g_{\mu\nu} + \frac 1 {\ell^2} g_{\mu\nu} = 0, \qquad \mbox{with }\mu,\nu=0,1,2,
\end{equation}
have 6 component equations.  Among these equations, there are 3 Bianchi identities
\begin{equation}
R_\mu^{~\nu}{}_{;\nu} -\frac 1 2 R_{;\mu} = 0.
\end{equation}
Therefore, to fix the solutions 3 coordinate conditions can be imposed.
We add these 3 conditions,
\begin{equation}
g_{01}=1,g_{11}=0,g_{12}=0,
\end{equation}
so the metric can be written as
\begin{equation}\label{eq:metric}
\od s^2 = g_{00}(\od x^0)^2 + 2 \od x^0 \od x^1 +2g_{02} \od x^0 \od x^2 +g_{22}(\od x^2)^2,
\end{equation}
and the inverse metric is
\begin{alignat}{1}
(g^{\mu\nu})=\left(\begin{array}{ccc}
              0 & 1 & 0 \smallskip\\
             1 & \dfrac{(g_{02})^{2}}{g_{22}}-g_{00} & -\dfrac{g_{02}}{g_{22}} \smallskip\\
              0 & -\dfrac{g_{02}}{g_{22}} & \dfrac{1}{g_{22}}
            \end{array}
          \right).
\end{alignat}
\omits{\begin{alignat}{1}
g:=det(g_{\mu\nu})=g_{22}.
\end{alignat}}
This is the metric in Bondi-like coordinates $(x^0,x^1,x^2)$.  It contains  only 3
independent components, which can be totally determined by the Einstein
field equations in 3-dimensional spacetime.

The most simple coframe fields contain only 3 independent variables which are equal to the number of the metric variables in \eqref{eq:metric},
and can be chosen as
\begin{alignat}{2}\label{eq:coframe-c1}
e^-=-\od x^0, \qquad e^+=e^{+}_{0}\od x^0 +\od x^1+e^{+}_{2}\od x^{2}, \qquad e^2=e^{2}_{2}\od x^2,
\end{alignat}
where $e^{+}_{0},\ e^{+}_{2},\ e^{2}_{2}$ are 3 arbitrary functions of the coordinates.
The metric and the coframe fields are related by
\begin{alignat}{1}
\od s^{2}=&\eta_{IJ}e^{I}\otimes e^{J}
=2e^{+}_0(\od x^0)^2+2\od x^0 \od x^1+2e^{+}_2\od x^0 \od x^2+(e^{2}_{2})^2(\od x^{2})^{2},
\end{alignat}
or
\begin{equation}
g_{00}=2e^{+}_{0}, \quad g_{01}=1,\quad g_{02}=e^{+}_{2}, \quad g_{11}=0, \quad g_{12}=0, \quad g_{22}=(e^{2}_{2})^{2}.
\end{equation}
where $I, J = -, +, 2$ are internal indices and $\eta_{IJ}$ is the metric of the internal space,
\begin{alignat}{1}
(\eta_{IJ})
=\left(     \begin{array}{ccc}
              0 & -1 & 0 \\
             -1 & 0 & 0 \\
              0 & 0 & 1 \\
            \end{array}
          \right).  \label{eq:internal-metric}
\end{alignat}
\omits{The inverse of the internal metric $\eta_{IJ}$ is
\begin{alignat}{1}
(\eta^{IJ})
=\left(     \begin{array}{ccc}
              0 & -1 & 0 \\
             -1 & 0 & 0 \\
              0 & 0 & 1 \\
            \end{array}
          \right).
\end{alignat}}
It is easy to check that
\begin{alignat}{6}
&g^{\mu\nu}e^{-}_{\mu}e^{-}_{\nu}=0, &\quad
g^{\mu\nu}e^{+}_{\mu}e^{+}_{\nu}=0, & \quad
g^{\mu\nu}e^{2}_{\mu}e^{2}_{\nu}=1, & \quad
&g^{\mu\nu}e^{-}_{\mu}e^{+}_{\nu}=-1, & \quad
g^{\mu\nu}e^{-}_{\mu}e^{2}_{\nu}=0, & \quad
g^{\mu\nu}e^{+}_{\mu}e^{2}_{\nu}=0,
\end{alignat}
so $e^{-}$, $e^{+}$ are null, and $e^{2}$ is spacelike.

Alternatively, the coframe fields can also be chosen as
\begin{alignat}{2}\label{eq:coframe-c2}
e^-=-\od x^0,\qquad e^+=e^{+}_{0}\od x^0 +\od x^1,\qquad e^2=e^{2}_{0}\od x^{0}+e^{2}_{2}\od x^2,
\end{alignat}
which have 3 independent variables as well. Under the coframe, the metric can be written as
\begin{alignat}{1}
\od s^{2}=&\eta_{IJ}e^{I}\otimes e^{J}=
[2e^{+}_0+(e^{2}_{0})^{2}](\od x^0)^2+2\od x^0 \od x^1+2e^{2}_{0}e^{2}_{2}\od x^{0}\od x^{2}+(e^{2}_{2})^2(\od x^{2})^2,
\end{alignat}
namely,
\begin{equation}
g_{00}=2e^{+}_{0}+(e^{2}_{0})^{2},\quad g_{01}=1,\quad g_{02}=e^{2}_{0}e^{2}_{2}, \quad g_{11}=0,
\quad g_{12}=0, \quad g_{22}=(e^{2}_{2})^{2}.
\end{equation}
In this case, $e^{-}$, $e^{+}$ are still null, and $e^{2}$ is still spacelike.

The 2 kinds of coframe fields with 3 independent variables are just the most simple choices.
A generic coframe which leaves the metric invariant may have more variables and can always
be obtained from the above simple choices by the following 3 kinds of basic gauge transformations
or their combinations (cf. \cite{BCG}):
\begin{alignat}{8}
\mbox{Boost transformation:}&\quad E^{-}=\frac{1}{\alpha}e^-,\quad E^{+}=\alpha e^+,\quad E^{2}=e^2, \label{eq:boost}\\
\mbox{Translation I:}&\quad E^{-}=e^-,\quad E^{+}=e^+-be^{2}+\frac{1}{2}b^{2}e^{-},\quad E^{2}=e^2-be^{-}, \label{eq:translation-}\\
\mbox{Translation II:}&\quad E^{-}=e^{-}-ce^{2}+\frac{1}{2}c^{2}e^{+},\quad E^{+}=e^+,\quad  E^{2}=e^2-ce^{+},\label{eq:translation+}
\end{alignat}
where $e^{I}$ may be either \eqref{eq:coframe-c1} or \eqref{eq:coframe-c2} and
$\alpha$, $b$ and $c$ are three independent arbitrary functions of coordinates.  It is easy to
check that
\begin{alignat}{1}
\od s^{2}=&\eta_{IJ}E^{I}\otimes E^{J}
=\eta_{IJ}e^{I}\otimes e^{J},
\end{alignat}
$E^{\pm}$ are lightlike and $E^{2}$ is spacelike.

The two choices of the most simple coframe fields are physically equivalent since they are  related to each other by a gauge transformation and can provide the same metric.
Although the relation between the metric and the coframe is clearer in the the choice \eqref{eq:coframe-c1}, the choice \eqref{eq:coframe-c2} is more commonly used such as in the discussion of BTZ spacetime.
Therefore, the following Hamiltonian analysis is based on \eqref{eq:coframe-c2}.

\subsection{Action}

In connection dynamics, the Palatini action of general relativity is commonly used, which is
equivalent to the Einstein-Hilbert action for non-degenerate cases.
The 3-dimensional Palatini action with a cosmological constant term can always be written as
\begin{alignat}{1}
S=\frac{1}{2}\int_{{\cal M}}f^{IJ}\wedge\sigma_{IJ}+\frac{\Lambda}{3!}\int_{\cal M}\epsilon_{IJK}e^{I}\wedge e^{J}\wedge e^{K}.
\end{alignat}
Here, $\Lambda=\ell^{-2}$ is the cosmological constant, $\epsilon_{IJK}$ is the Levi-Civita symbol,
\begin{alignat}{1}
\sigma_{IJ}=\epsilon_{IJK}e^{K}
\end{alignat}
is the ``area'' element in the 3-dimensional spacetime,
$f^{IJ}$ is the curvature tensor of the connection $\omega^{IJ}$,
\begin{alignat}{1}
f^{IJ}=\od\omega^{IJ}+\eta_{KL}\omega^{IK}\wedge\omega^{LJ},
\end{alignat}
where
$\omega^{IJ}$ are the $\frak{so}(1,2)$-connection 1-forms, which satisfy torsion-free condition,
\begin{alignat}{1}
\od e^{I}+\omega^{IJ}\wedge e^{K}\eta_{JK}=0.\label{2.66}
\end{alignat}
Therefore, the action can be expanded as
\begin{alignat}{1}
S
=&\int_{\cal M}(-f^{-2}\wedge e^{+}+f^{-+}\wedge e^{2}+f^{+2}\wedge e^{-}+\Lambda e^{-}\wedge e^{+}\wedge e^{2}).
\end{alignat}
For the coframe fields (\ref{eq:coframe-c2}), the action becomes
\begin{alignat}{1}
S=&\int_{\cal M}\left [f^{-2}_{02}-e^{+}_{0}f^{-2}_{12}+e^{2}_{0}f^{-+}_{12}-f^{+2}_{12} +e^{2}_{2}(f^{-+}_{01}
-\Lambda)\right]\od^3 x =\int_{\cal M}\mathcal{L}(e^{I}_{\mu},\omega^{JK}_{\nu})\od^3x.\label{2.69}
\end{alignat}

Under the 3 kinds of basic gauge transformations \eqref{eq:boost}, \eqref{eq:translation-}, and
\eqref{eq:translation+}, the torsion-free connection become
\begin{alignat}{6}
\mbox{Boost:}&\quad\Omega^{-+}=\omega^{-+}-\od\ln\alpha,\quad \Omega^{-2}=\frac{1}{\alpha}\omega^{-2},
\quad\Omega^{+2}=\alpha\omega^{+2} \\
\mbox{Translation I:}&\quad \Omega^{-+}=\omega^{-+}-b\omega^{-2},\quad\Omega^{-2}=\omega^{-2},
\quad\Omega^{+2}=\omega^{+2}+b\omega^{-+}+\od b-\frac{1}{2}b^{2}\omega^{-2},\\
\mbox{Translation II:}&\quad\Omega^{-+}=\omega^{-+}+c\omega^{+2},
\quad \Omega^{-2}=\omega^{-2}-c\omega^{-+}+\od c-\frac{1}{2}c^{2}\omega^{+2},
\quad \Omega^{+2}=\omega^{+2},
\end{alignat}
respectively, where $\Omega^{IJ}$ satisfy
\begin{alignat}{1}
\od E^{I}+\Omega^{IJ}\wedge E^{K}\eta_{JK}=0. \label{2.70}
\end{alignat}
The field strengths change as
\begin{alignat}{8}
\mbox{Boost:}&\quad F^{-+}=f^{-+}, \quad F^{-2}=\frac{1}{\alpha}f^{-2},\quad F^{+2}=\alpha f^{+2},\\
\mbox{Translation I:}&\quad F^{-+}=f^{-+}-bf^{-2}, \quad F^{-2}=f^{-2},\quad F^{+2}=f^{+2}+bf^{-+}-\frac{1}{2}b^{2}f^{-2},\\
\mbox{Translation II:}&\quad F^{-+}=f^{-+}+cf^{+2},\quad F^{-2}=f^{-2}-cf^{-+}-\frac{1}{2}c^{2}f^{+2},\quad F^{+2}=f^{+2},
\end{alignat}
respectively.  Then, the action
\begin{alignat}{1}
S=&\int_{M}\frac{1}{2}\epsilon_{IJK}F^{IJ}\wedge E^{K}+\frac{\Lambda}{3!}\epsilon_{IJK}E^{I}\wedge E^{J}\wedge E^{K}
=\int_{M}\frac{1}{2}\epsilon_{IJK}f^{IJ}\wedge e^{K}+\frac{\Lambda}{3!}\epsilon_{IJK}e^{I}\wedge e^{J}\wedge e^{K},
\end{alignat}
is invariant under the above 3 transformations.
Therefore, in the following analysis, only the most simple coframe fields are considered.

\subsection{Hamiltonian}
For the above action (\ref{2.69}), the Lagrangian is
\begin{alignat}{1}
L=&\int_{\cal S}\mathcal{L}\od^2 x
=\int_{\cal S}(e^{2}_{2}f^{-+}_{01}+e^{2}_{0}f^{-+}_{12}+f^{-2}_{02}-e^{+}_{0}f^{-2}_{12}-f^{+2}_{12}
-\Lambda e^{2}_{2})\od^2x \notag  \\
=&\int_{\cal S}[(\omega^{-2}_{2,0}-\omega^{-2}_{0,2}+\omega^{-+}_{2}\omega^{-2}_{0}
-\omega^{-+}_{0}\omega^{-2}_{2}) -e^{+}_{0}(\omega^{-2}_{2,1}-\omega^{-2}_{1,2}
+\omega^{-+}_{2}\omega^{-2}_{1}-\omega^{-+}_{1}\omega^{-2}_{2}) \nonumber \\
&\qquad +e^{2}_{0}(\omega^{-+}_{2,1}-\omega^{-+}_{1,2}+\omega^{-2}_{2}\omega^{+2}_{1}
-\omega^{-2}_{1}\omega^{+2}_{2})-(\omega^{+2}_{2,1}-\omega^{+2}_{1,2}
+\omega^{-+}_{1}\omega^{+2}_{2}-\omega^{-+}_{2}\omega^{+2}_{1})\nonumber \\
&\qquad +e^{2}_{2}(\omega^{-+}_{1,0}-\omega^{-+}_{0,1}+\omega^{-2}_{1}\omega^{+2}_{0}
-\omega^{-2}_{0}\omega^{+2}_{1}) -\Lambda e^{2}_{2}]\od^2x, \label{eq:Lagrangian}
\end{alignat}
where ${\cal S}$ denotes the hypersurface 
at constant $x^{0}$.  In \eqref{eq:Lagrangian},
$e^I_\mu$ and $\omega^{IJ}_{\mu}$ are treated as independent canonical configuration variables at the beginning,
which are denoted by $Q^\beta$ in a unified way.
It will be seen that the torsion-free conditions will come out as secondary constraints and
equations of motion.  Therefore, there is no need to add the torsion-free conditions as
primary constraints in the Lagrangian.
The canonical momenta $P_\beta$ conjugate to $Q^\beta$ are defined by
\begin{alignat}{1}
P_{\beta}:=\frac{\delta L}{\delta Q^{\beta}_{,0}}.
\end{alignat}
$\pi^{\mu}_{I}$ and $\pi^{\mu}_{IJ}$ are the canonical momenta conjugate to $e^{I}_{\mu}$
and $\omega^{IJ}_{\mu}$, respectively.  Namely,
\begin{alignat}{1}
&\pi^{\mu}_{I}:=\frac{\delta L}{\delta\dot{e}^{I}_{\mu}}, \\
&\pi^{\mu}_{IJ}:=\frac{\delta L}{\delta\dot{\omega}^{IJ}_{\mu}}. \label{eq:cm:piIJ}
\end{alignat}
The conditions of the 12 conjugate momenta are treated as primary constraints
\begin{alignat}{2}
\phi^{0}_{+}:=&\pi^{0}_{+}=0, \quad \phi^{0}_{2}:=\pi^{0}_{2}=0, \quad  \phi^{2}_{2}:=\pi^{2}_{2}=0,
\nonumber \\
\phi^{0}_{-+}:=&\pi^{0}_{-+}=0, \quad \phi^{1}_{-+}:=\pi^{1}_{-+}-e^{2}_{2}=0, \quad \phi^{2}_{-+}:=\pi^{2}_{-+}=0,
\nonumber \\
\phi^{0}_{-2}:=&\pi^{0}_{-2}=0, \quad \phi^{1}_{-2}:=\pi^{1}_{-2}=0, \quad \phi^{2}_{-2}:=\pi^{2}_{-2}-1=0, 
\nonumber \\
\phi^{0}_{+2}:=&\pi^{0}_{+2}=0, \quad \phi^{1}_{+2}:=\pi^{1}_{+2}=0, \quad \phi^{2}_{+2}:=\pi^{2}_{+2}=0,
\end{alignat}
denoted by $\phi_\xi$ in brief.

The free form of Hamiltonian is given by the Legendre transformation,
\begin{alignat}{1}
H_{c}=&\int_{\cal S}\mathcal{H}_{c}\od^2 x
=\int_{\cal S}(P_{\beta}\dot{Q}^{\beta}-\mathcal{L})\od^2x \nonumber \\
=&\int_{\cal S}[e^{2}_{2}(\omega^{-+}_{0,1}-\omega^{-2}_{1}\omega^{+2}_{0}+\omega^{-2}_{0}\omega^{+2}_{1})
+(\omega^{-2}_{0,2}-\omega^{-+}_{2}\omega^{-2}_{0}+\omega^{-+}_{0}\omega^{-2}_{2})
-e^{2}_{0}f^{-+}_{12}+e^{+}_{0}f^{-2}_{12}+f^{+2}_{12}+\Lambda e^{2}_{2}]\od^2x.
\end{alignat}
Obviously, the system is a constrained one.
The 
primary constraints should be added into the Hamiltonian \cite{Dirac} to get a consistent theory,
\begin{alignat}{1}
H_T :=&H_{c}+H_{1}
=\int_{\cal S}(\mathcal{H}_{c}+\lambda^{\xi}\phi_{\xi})\od^2x \nonumber \\
=&\int_{\cal S}\od^2 x \left [ e^{2}_{2}(\omega^{-+}_{0,1}-\omega^{-2}_{1}\omega^{+2}_{0}
+\omega^{-2}_{0}\omega^{+2}_{1})\right.
+(\omega^{-2}_{0,2}-\omega^{-+}_{2}\omega^{-2}_{0}+\omega^{-+}_{0}\omega^{-2}_{2})
-e^{2}_{0}f^{-+}_{12}+e^{+}_{0}f^{-2}_{12}+f^{+2}_{12}+\Lambda e^{2}_{2}
\nonumber \\
&\qquad+\lambda^{+}_{0}\pi^{0}_{+}
+\lambda^{2}_{0}\pi^{0}_{2}+\lambda^{2}_{2}\pi^{2}_{2}+\lambda^{-+}_{0}\pi^{0}_{-+}+\lambda^{-+}_{1}(\pi^{1}_{-+}-e^{2}_{2})
+\lambda^{-+}_{2}\pi^{2}_{-+}+\lambda^{-2}_{0}\pi^{0}_{-2}+\lambda^{-2}_{1}\pi^{1}_{-2}
\nonumber \\
&\qquad \left . +\lambda^{-2}_{2}(\pi^{2}_{-2}-1) +\lambda^{+2}_{0}\pi^{0}_{+2}
+\lambda^{+2}_{1}\pi^{1}_{+2}+\lambda^{+2}_{2}\pi^{2}_{+2}\right ],
\end{alignat}
where $\lambda^\xi$ are Lagrangian multipliers which should be determined by the following analysis.

\section{Consistency Conditions of the Constraints}

All constraints $\Phi_R$, including primary constraints $\phi_\xi$ and possible secondary constraints, denoted by $\psi_n$, should satisfy the consistency conditions on the constraint surface in phase space,
\begin{alignat}{1}
\{H_T ,\Phi_R\}=\int_{\cal S}\{\mathcal{H}_{c},\Phi_R\}\od^2x+\int_{\cal S}\lambda^{\xi}\{\phi_{\xi},\Phi_{R}\}\od^2x\approx0,
\end{alignat}
where $\{\quad ,\quad \}$ is the Poisson bracket, $``\approx"$ means ``equal to" on the constraint surface.
\subsection{Consistency Conditions of the Primary Constraints}
\begin{alignat}{2}
&\{H_T ,\pi^{0}_{+}\}=\frac{\delta H_T }{\delta e^{+}_{0}}=f^{-2}_{12}
=\omega^{-2}_{2,1}-\omega^{-2}_{1,2}-\omega^{-+}_{1}\omega^{-2}_{2}
+\omega^{-+}_{2}\omega^{-2}_{1}\approx 0,\label{22} \\
&\{H_T ,\pi^{0}_{2}\}=\frac{\delta H_T }{\delta e^{2}_{0}}=-f^{-+}_{12}
=\omega^{-+}_{2,1}-\omega^{-+}_{1,2}-\omega^{-2}_{1}\omega^{+2}_{2}
+\omega^{-2}_{2}\omega^{+2}_{1}\approx 0,\label{24} \\
& \{H_T ,\pi^{2}_{2}\}=\frac{\delta H_T }{\delta e^{2}_{2}}
=\omega^{-+}_{0,1}-\omega^{-2}_{1}\omega^{+2}_{0}+\omega^{-2}_{0}\omega^{+2}_{1}
-\lambda^{-+}_{1}+\Lambda\approx0,\label{25} \\
& \{H_T ,\pi^{0}_{-+}\}=\frac{\delta H_T }{\delta\omega^{-+}_{0}}
=-e^{2}_{2,1}+\omega^{-2}_{2}\approx 0,\label{26}\\
& \{H_T ,\pi^{1}_{-+}-e^{2}_{2}\}=\frac{\delta H_T }{\delta\omega^{-+}_{1}}+\frac{\delta H_T }{\delta\pi^{2}_{2}} =-e^{2}_{0,2}-\omega^{-2}_{2} e^{+}_{0}
+\omega^{+2}_{2}+\lambda^{2}_{2}\approx0,\label{27} \\
& \{H_T ,\pi^{2}_{-+}\}=\frac{\delta H_T }{\delta\omega^{-+}_{2}}
=e^{2}_{0,1}-\omega^{-2}_{0}-\omega^{+2}_{1}+\omega^{-2}_{1}e^{+}_{0}\approx 0,\label{28} \\
& \{H_T ,\pi^{0}_{-2}\}=\frac{\delta H_T }{\delta\omega^{-2}_{0}}
=\omega^{+2}_{1}e^{2}_{2}+\omega^{-+}_{1}e^{+}_{2}-\omega^{-+}_{2}\approx 0,\label{29}\\
&\{H_T ,\pi^{1}_{-2}\}=\frac{\delta H_T }{\delta\omega^{-2}_{1}}
=e^{+}_{0,2}-\omega^{+2}_{0}e^{2}_{2}+\omega^{-+}_{2}e^{+}_{0}+\omega^{+2}_{2}e^{2}_{0}\approx0, \label{30} \\
&\{H_T ,\pi^{2}_{-2}-1\}=\frac{\delta H_T }{\delta\omega^{-2}_{2}}
=-e^{+}_{0,1}+\omega^{-+}_{0}-\omega^{-+}_{1}e^{+}_{0}-\omega^{+2}_{1}e^{2}_{0}\approx 0,\label{31} \\
&\{H_T ,\pi^{0}_{+2}\}=\frac{\delta H_T }{\delta\omega^{+2}_{0}}=-\omega^{-2}_{1}e^{2}_{2}\approx 0,\label{32} \\
&\{H_T ,\pi^{1}_{+2}\}=\frac{\delta H_T }{\delta\omega^{+2}_{1}}
=-\omega^{-+}_{2}+\omega^{-2}_{0}e^{2}_{2}-\omega^{-2}_{2}e^{2}_{0}\approx0,\label{33} \\
&\{H_T ,\pi^{2}_{+2}\}=\frac{\delta H_T }{\delta\omega^{+2}_{2}}=\omega^{-2}_{1}e^{2}_{0}+\omega^{-+}_{1}\approx0.\label{34}
\end{alignat}
Eq. ($\ref{32}$) requires
\begin{alignat}{1}
\psi_1:=\omega^{-2}_{1}\approx 0,\label{38}
\end{alignat}
because $e^{2}_{2}$ is not equal to 0.  Then, Eq. ($\ref{34}$) reduces to
\begin{alignat}{1}
\psi_2:=\omega^{-+}_{1}\approx 0.\label{eq:psi2}
\end{alignat}
Eqs. ($\ref{22}$) and ($\ref{24}$) are simplified to
\begin{alignat}{2}
& \psi_3:=\omega^{-2}_{2,1}\approx0, \label{eq:psi3}\\
& \psi_4:=\omega^{-+}_{2,1}+\omega^{-2}_{2}\omega^{+2}_{1}\approx0.\label{4}
\end{alignat}
Eqs.(\ref{25}) and (\ref{27}) determine 2 Lagrange multipliers directly,
\begin{alignat}{2}
&\lambda^{-+}_{1}\approx\omega^{-+}_{0,1}+\omega^{-2}_{0}\omega^{+2}_{1}+\Lambda=:\Lambda^{-+}_{1},\label{3.17}\\
&\lambda^{2}_{2}\approx e^{2}_{0,2}+\omega^{-2}_{2}e^{+}_{0}-\omega^{+2}_{2}=:\Lambda^{2}_{2}, \label{eq:lambda2_2}
\end{alignat}
where the capital symbols $\Lambda^{IJ}_{\mu}$ and $\Lambda^{I}_{\mu}$ represent the solutions of the multipliers.
Eqs.(\ref{26}), (\ref{28}), (\ref{29}), (\ref{30}), (\ref{31}), and (\ref{33}) can be written as
\begin{alignat}{2}
&\psi_5:=e^{2}_{2,1}-\omega^{-2}_{2}\approx 0, \label{eq:psi5}\\
&\psi_6:=e^{2}_{0,1}-\omega^{-2}_{0}-\omega^{+2}_{1}\approx0, \label{eq:psi6}\\
&\psi_7:=\omega^{-+}_{2}-\omega^{+2}_{1}e^{2}_{2}\approx0, \label{eq:psi7}\\
&\psi_8:=e^{+}_{0,2}-\omega^{+2}_{0}e^{2}_{2}+\omega^{-+}_{2}e^{+}_{0}+\omega^{+2}_{2}e^{2}_{0}
\approx0,\label{eq:psi8}\\
&\psi_9:=e^{+}_{0,1}-\omega^{-+}_{0}+\omega^{+2}_{1}e^{2}_{0}\approx0,\label{eq:psi9}\\
&\psi_{10}:=\omega^{-+}_{2}-\omega^{-2}_{0}e^{2}_{2}+\omega^{-2}_{2}e^{2}_{0}\approx0,\label{eq:psi10}
\end{alignat}
which are exactly the torsion-free conditions, because the torsion-free condition reads
\begin{alignat}{1}
De^{I}=\textrm{d}e^{I}+\omega^{IJ}\wedge e^{K}\eta_{JK}=0,
\end{alignat}
and its component equations are
\begin{alignat}{1}
2D_{[\mu}e^{I}_{\nu]}=-e^{I}_{\mu,\nu}+e^{I}_{\nu,\mu}+\omega^{IJ}_{\mu}e^{K}_{\nu}\eta_{JK}-\omega^{IJ}_{\nu}e^{K}_{\mu}\eta_{JK}=0.
\end{alignat}
In fact, Eq.\eqref{32} and Eq.\eqref{34} are also the torsion-free conditions.

In summary, the consistency conditions for the primary constraints provide 10 secondary constraints
and the explicit expressions for 2 Lagrange multipliers.

\subsection{Consistency Conditions of the Secondary Constraints}
The consistency conditions for the secondary constraints are
\begin{alignat}{2}
\{H_T ,\psi_1\}=&\{H_T ,\omega^{-2}_{1}\}=-\lambda^{-2}_{1}\approx 0, \label{eq:cc-psi1}\\
\{H_T ,\psi_2\}=&\{H_T ,\omega^{-+}_{1}\}=-\lambda^{-+}_{1}\approx 0, \label{eq:cc-psi2}\\
\{H_T ,\psi_3\}=&\{H_T ,\omega^{-2}_{2,1}\}=-\lambda^{-2}_{2,1}\approx0,\label{eq:cc-psi3}\\
\{H_T ,\psi_4\}=&\{H_T ,\omega^{-+}_{2,1}+\omega^{-2}_{2}\omega^{+2}_{1}\}
=-\lambda^{-+}_{2,1}-\lambda^{-2}_{2}\omega^{+2}_{1}-\lambda^{+2}_{1}\omega^{-2}_{2}\approx0,\label{eq:cc-psi4}\\
\{H_T ,\psi_5\}=&\{H_T ,e^{2}_{2,1}-\omega^{-2}_{2}\}=-\lambda^{2}_{2,1}+\lambda^{-2}_{2}\approx0, \label{eq:cc-psi5}\\
\{H_T ,\psi_6\}=&\{H_T ,e^{2}_{0,1}-\omega^{-2}_{0}-\omega^{+2}_{1}\}
=-\lambda^{2}_{0,1}+\lambda^{-2}_{0}+\lambda^{+2}_{1}\approx0, \label{eq:cc-psi6}\\
\{H_T ,\psi_7\}=&\{H_T ,\omega^{-+}_{2}-\omega^{+2}_{1}e^{2}_{2}\}
=-\lambda^{-+}_{2}+\lambda^{+2}_{1}e^{2}_{2}+\omega^{+2}_{1}\lambda^{2}_{2}\approx0,
\label{eq:cc-psi7} \\
\{H_T ,\psi_8\}=&\{H_T ,e^{+}_{0,2}-\omega^{+2}_{0}e^{2}_{2}+\omega^{-+}_{2}e^{+}_{0}
+\omega^{+2}_{2}e^{2}_{0}\} \nonumber \\
=&-\lambda^{+}_{0,2}+\lambda^{+2}_{0}e^{2}_{2}+\omega^{+2}_{0}\lambda^{2}_{2}
-\lambda^{-+}_{2}e^{+}_{0}-\omega^{-+}_{2}\lambda^{+}_{0}-\lambda^{+2}_{2}e^{2}_{0}
-\omega^{+2}_{2}\lambda^{2}_{0}\approx0,\label{3.34}\\
\{H_T ,\psi_{9}\}=&\{H,e^{+}_{0,1}-\omega^{-+}_{0}+\omega^{+2}_{1}e^{2}_{0}\}
=-\lambda^{+}_{0,1}+\lambda^{-+}_{0}-\lambda^{+2}_{1}e^{2}_{0}-\omega^{+2}_{1}\lambda^{2}_{0}
\approx0,\label{3.35}\\
\{H_T ,\psi_{10}\}=&\{H_T ,\omega^{-+}_{2}-\omega^{-2}_{0}e^{2}_{2}+\omega^{-2}_{2}e^{2}_{0}\}
=-\lambda^{-+}_{2}+\lambda^{-2}_{0}e^{2}_{2}+\omega^{-2}_{0}\lambda^{2}_{2}
-\lambda^{-2}_{2}e^{2}_{0}-\omega^{-2}_{2}\lambda^{2}_{0}\approx0.\label{eq:cc-psi10}
\end{alignat}
The combination of (\ref{3.17}) and (\ref{eq:cc-psi2}) leads to a new constraint
\begin{alignat}{2}\label{eq:psi11}
\psi_{11}:=\omega^{-+}_{0,1}+\omega^{-2}_{0}\omega^{+2}_{1}+\Lambda\approx0.
\end{alignat}
(\ref{eq:cc-psi5}) results in
\begin{alignat}{2}
\lambda^{-2}_{2}\approx\lambda^{2}_{2,1}\approx\Lambda^{2}_{2,1}=:\Lambda^{-2}_{2}. \label{eq:lambda-2_2}
\end{alignat}
Then, (\ref{eq:cc-psi3}) provides a new constraint
\begin{alignat}{2}\label{eq:psi12}
\psi_{12}:=\Lambda^{-2}_{2,1}=\Lambda^{2}_{2,11}\approx0.
\end{alignat}
\omits{The last equality can be integrated out
\begin{alignat}{2}
\Lambda^{2}_{2}\approx C_{5}x^{1}+C_{6},  \qquad\mbox{with\quad}
C_{5,1}=C_{6,1}=0.
\end{alignat}
In fact, $C_5$ must vanish because a constraint is a relation among canonical quantities.  Hence,
the constraint \eqref{eq:psi12} becomes then
\begin{alignat}{2}\label{eq:psi12'}
\psi'_{12}:=\Lambda^{-2}_{2}=\Lambda^{2}_{2,1}\approx0.
\end{alignat}}

By use of \eqref{eq:psi5}, \eqref{eq:lambda2_2}, and \eqref{eq:lambda-2_2},
one can obtain from (\ref{eq:cc-psi4}) and (\ref{eq:cc-psi7})
\begin{alignat}{2}
\lambda^{-+}_{2}&\approx\frac{1}{e^{2}_{2}}\int\omega^{+2}_{1}(\omega^{-2}_{2}\Lambda^{2}_{2}
-\Lambda^{-2}_{2}e^{2}_{2})\od x^{1}+\frac{1}{e^{2}_{2}}C_{1}=:\Lambda^{-+}_{2}, \\
\lambda^{+2}_{1}&\approx\frac{1}{e^{2}_{2}}(\Lambda^{-+}_{2}-\omega^{+2}_{1}\Lambda^{2}_{2})=:\Lambda^{+2}_{1},
\end{alignat}
where $C_{1}$ is an integral constant with $C_{1,1}=0$.  From (\ref{eq:cc-psi10}) and (\ref{eq:cc-psi6}), finally, one can acquire
\begin{alignat}{2}
\lambda^{2}_{0}&\approx e^{2}_{2}\int\frac{1}{(e^{2}_{2})^{2}}
(\Lambda^{-+}_{2}+\Lambda^{+2}_{1}e^{2}_{2}-\omega^{-2}_{0}\Lambda^{2}_{2}
+\Lambda^{-2}_{2}e^{2}_{0})\od x^{1}+e^{2}_{2}C_{2}=:\Lambda^{2}_{0},\\
\lambda^{-2}_{0}&\approx(\Lambda^{2}_{0,1}-\Lambda^{+2}_{1})=:\Lambda^{-2}_{0},
\end{alignat}
where $C_{2}$ is another integral constant with $C_{2,1}=0$.

In summary, the consistency conditions for the secondary constraints determine the other 6
Lagrange multipliers, set 2 new secondary constraints, and present 2 conditions about multipliers.

\subsection{Consistency Conditions of the Further Secondary Constraints}

The consistency condition of the first further secondary constraint \eqref{eq:psi11} is
\begin{alignat}{1}
\{H_T ,\psi_{11}\}=\{H_T ,\omega^{-+}_{0,1}+\omega^{-2}_{0}\omega^{+2}_{1}+\Lambda\}
=-\lambda^{-+}_{0,1}-\lambda^{-2}_{0}\omega^{+2}_{1}-\omega^{-2}_{0}\lambda^{+2}_{1}\approx0.
\end{alignat}
With the help of the above solved multipliers, one gets
\begin{alignat}{1}
\lambda^{-+}_{0}\approx-\int(\Lambda^{-2}_{0}\omega^{+2}_{1}+\omega^{-2}_{0}\Lambda^{+2}_{1})\od
x^{1}+C_{3}=:\Lambda^{-+}_{0},
\end{alignat}
where $C_{3,1}=0$.  Then, (\ref{3.35}) becomes
\begin{alignat}{1}
\lambda^{+}_{0,1}-\Lambda^{-+}_{0}+\Lambda^{+2}_{1}e^{2}_{0}+\omega^{+2}_{1}\Lambda^{2}_{0}\approx0.
\end{alignat}
So,
\begin{alignat}{1}
\lambda^{+}_{0}\approx\int(\Lambda^{-+}_{0}-\Lambda^{+2}_{1}e^{2}_{0}-\omega^{+2}_{1}\Lambda^{2}_{0})
\od x^{1}+C_{4}=:\Lambda^{+}_{0}
\end{alignat}
with $C_{4,1}=0$.

The second further secondary constraint \eqref{eq:psi12} can be written explicitly as
\begin{alignat}{2}
\psi_{12}\approx (e^{2}_{0,2}+\omega^{-2}_{2}e^{+}_{0}-\omega^{+2}_{2})_{,11}\approx0.\label{3.50}
\end{alignat}
Its consistency condition is
\begin{alignat}{1}
\{H_T ,\psi_{12}\}\approx\{H_T ,(e^{2}_{0,2}+\omega^{-2}_{2}e^{+}_{0}-\omega^{+2}_{2})_{,11}\}
=-\lambda^{2}_{0,211}-\lambda^{-2}_{2}e^{+}_{0,11}-\omega^{-2}_{2}\lambda^{+}_{0,11}
+\lambda^{+2}_{2,11}\approx0.
\end{alignat}
So,
\begin{alignat}{1} \label{eq:lambda+2_2}
(\lambda^{+2}_{2}-\Lambda^{2}_{0,2}-\Lambda^{-2}_{2}e^{+}_{0}
-\omega^{-2}_{2}\Lambda^{+}_{0})_{,11}\approx0,
\end{alignat}
whose solution for $\lambda^{+2}_{2}$ is denoted by $\Lambda^{+2}_{2}$.
Finally, from (\ref{3.34}) one can solve the multiplier $\lambda^{+2}_0$,
\begin{alignat}{1}\label{eq:lambda+2_0}
\lambda^{+2}_{0}\approx\frac{1}{e^{2}_{2}}(\Lambda^{+}_{0,2}-\omega^{+2}_{0}\Lambda^{2}_{2}+\Lambda^{-+}_{2}e^{+}_{0}+\omega^{-+}_{2}\Lambda^{+}_{0}
+\Lambda^{+2}_{2}e^{2}_{0}+\omega^{+2}_{2}\Lambda^{2}_{0})=:\Lambda^{+2}_{0}.
\end{alignat}

\omits{The second further secondary constraint \eqref{eq:psi12'} can be written explicitly as
\begin{alignat}{2}
\psi'_{12} \approx (e^{2}_{0,2}+\omega^{-2}_{2}e^{+}_{0}-\omega^{+2}_{2})_{,1}\approx0.\label{3.50}
\end{alignat}
Its consistency condition is
\begin{alignat}{1}
\{H_T ,\psi'_{12}\}\approx\{H_T ,(e^{2}_{0,2}+\omega^{-2}_{2}e^{+}_{0}-\omega^{+2}_{2})_{,1}\}
=\lambda^{2}_{0,21}+\lambda^{-2}_{2}e^{+}_{0,1}+\omega^{-2}_{2}\lambda^{+}_{0,1}
-\lambda^{+2}_{2,1}\approx0.
\end{alignat}
So,
\begin{alignat}{1} \label{eq:lambda+2_2}
(\lambda^{+2}_{2}-\Lambda^{2}_{0,2}-\Lambda^{-2}_{2}e^{+}_{0}
-\omega^{-2}_{2}\Lambda^{+}_{0})_{,1}\approx0,
\end{alignat}
whose solution for $\lambda^{+2}_{2}$ is denoted by $\Lambda^{+2}_{2}$.
Finally, from (\ref{3.34}) one can solve the multiplier $\lambda^{+2}_0$,
\begin{alignat}{1}\label{eq:lambda+2_0}
\lambda^{+2}_{0}\approx\frac{1}{e^{2}_{2}}(\Lambda^{+}_{0,2}-\omega^{+2}_{0}\Lambda^{2}_{2}+\Lambda^{-+}_{2}e^{+}_{0}+\omega^{-+}_{2}\Lambda^{+}_{0}
+\Lambda^{+2}_{2}e^{2}_{0}+\omega^{+2}_{2}\Lambda^{2}_{0})=:\Lambda^{+2}_{0}.
\end{alignat}}
In fact, \eqref{eq:psi12} can be integrated out
\begin{alignat}{2}
\Lambda^{2}_{2}\approx C_{5}x^{1}+C_{6},  \qquad\mbox{with\quad}
C_{5,1}=C_{6,1}=0.
\end{alignat}
If $\Lambda^{2}_{2}=0$ is set at the 2 boundaries (denoted by `down' and `up', respectively) of the coordinate $x^{1}$, then one gets
\begin{alignat}{2}
&0=\Lambda^{2}_{2}\mid_{\rm down}\, \approx C_{5}x^{1}\mid_{\rm down}+C_{6},\\
&0=\Lambda^{2}_{2}\mid_{\rm up}\, \approx C_{5}x^{1}\mid_{\rm up}+C_{6},
\end{alignat}
which results in $C_{5}\approx C_{6}\approx 0$.
In this case, \eqref{eq:psi12} reduces to
\begin{alignat}{2}
\psi'_{12}:=\Lambda^{2}_{2}=e^{2}_{0,2}+\omega^{-2}_{2}e^{+}_{0}-\omega^{+2}_{2}\approx0.
\end{alignat}
Its consistency condition is
\begin{alignat}{2}
\{H_T ,\psi'_{12}\}=\{H_T ,e^{2}_{0,2}+\omega^{-2}_{2}e^{+}_{0}-\omega^{+2}_{2}\}
=-\lambda^{2}_{0,2}-\lambda^{-2}_{2}e^{+}_{0}-\omega^{-2}_{2}\lambda^{+}_{0}+\lambda^{+2}_{2}\approx0,
\end{alignat}
and thus,
\begin{alignat}{2}
\lambda^{+2}_{2}
\approx\Lambda^{2}_{0,2}-\Lambda^{-2}_{2}e^{+}_{0}-\omega^{-2}_{2}\Lambda^{+}_{0}
=:\tilde{\Lambda}^{+2}_{2}.
\end{alignat}
This is a special case for the solution of \eqref{eq:lambda+2_2}.
In the special case, \eqref{eq:lambda+2_0} reduces to
\begin{alignat}{1}
\lambda^{+2}_{0}\approx\frac{1}{e^{2}_{2}}(\Lambda^{+}_{0,2}-\omega^{+2}_{0}\Lambda^{2}_{2}+\Lambda^{-+}_{2}e^{+}_{0}+\omega^{-+}_{2}\Lambda^{+}_{0}
+\tilde{\Lambda}^{+2}_{2}e^{2}_{0}+\omega^{+2}_{2}\Lambda^{2}_{0})=:\tilde{\Lambda}^{+2}_{0}.
\end{alignat}

Up to now, all constraints (12 primary and 12 secondary constraints) are obtained and consistent in the evolution direction.
All 12 Lagrangian multipliers are determined.

\subsection{Degree of Freedom}

The coframe has 3 independent variables, and the connection coefficients have 9 variables.
The 12 variables and their conjugate momenta span a 24-dimensional phase space.  It is easy to see that the 24 constraints, including 12 primary constraints and 12 secondary constraints, are all second class.
Therefore, there is no local physical degree of freedom left in 3-dimensional gravitational system as expected.

\section{Equations of Motion}

The equations of motion are given by the Hamiltonian equations:
\begin{alignat}{1}
&\dot{e}^{+}_{0}=\{e^{+}_{0},H_{T}\}=\frac{\delta H_{T}}{\delta\pi^{0}_{+}}=\lambda^{+}_{0}\approx\Lambda^{+}_{0},\\
&\dot{e}^{2}_{0}=\{e^{2}_{0},H_{T}\}=\frac{\delta H_{T}}{\delta\pi^{0}_{2}}=\lambda^{2}_{0}\approx\Lambda^{2}_{0},\\
&\dot{e}^{2}_{2}=\{e^{2}_{2},H_{T}\}=\frac{\delta H_{T}}{\delta\pi^{2}_{2}}=\lambda^{2}_{2}\approx\Lambda^{2}_{2},\label{5.3}\\
&\dot{\omega}^{-+}_{0}=\{\omega^{-+}_{0},H_{T}\}=\frac{\delta H_{T}}{\delta\pi^{0}_{-+}}=\lambda^{-+}_{0}\approx\Lambda^{-+}_{0},\\
&\dot{\omega}^{-+}_{1}=\{\omega^{-+}_{1},H_{T}\}=\frac{\delta H_{T}}{\delta\pi^{1}_{-+}}=\lambda^{-+}_{1}\approx0,\\&
\dot{\omega}^{-+}_{2}=\{\omega^{-+}_{2},H_{T}\}=\frac{\delta H_{T}}{\delta\pi^{2}_{-+}}=\lambda^{-+}_{2}\approx\Lambda^{-+}_{2},\\
&\dot{\omega}^{-2}_{0}=\{\omega^{-2}_{0},H_{T}\}=\frac{\delta H_{T}}{\delta\pi^{0}_{-2}}=\lambda^{-2}_{0}\approx\Lambda^{-2}_{0},\\ 
&\dot{\omega}^{-2}_{1}=\{\omega^{-2}_{1},H_{T}\}=\frac{\delta H_{T}}{\delta\pi^{1}_{-2}}=\lambda^{-2}_{1}\approx0,\\ 
&\dot{\omega}^{-2}_{2}=\{\omega^{-2}_{2},H_{T}\}=\frac{\delta H_{T}}{\delta\pi^{2}_{-2}}=\lambda^{-2}_{2}\approx\Lambda^{-2}_{2},\\ 
&\dot{\omega}^{+2}_{0}=\{\omega^{+2}_{0},H_{T}\}=\frac{\delta H_{T}}{\delta\pi^{0}_{+2}}=\lambda^{+2}_{0}\approx\Lambda^{+2}_{0},\\ 
&\dot{\omega}^{+2}_{1}=\{\omega^{+2}_{1},H_{T}\}=\frac{\delta H_{T}}{\delta\pi^{1}_{+2}}=\lambda^{+2}_{1}\approx\Lambda^{+2}_{1},\\ 
&\dot{\omega}^{+2}_{2}=\{\omega^{+2}_{2},H_{T}\}=\frac{\delta H_{T}}{\delta\pi^{2}_{+2}}=\lambda^{+2}_{2}\approx\Lambda^{+2}_{2},\\
&\dot{\pi}^{0}_{+}=\{\pi^{0}_{+},H_{T}\}=-f^{-2}_{12}\approx0,\\
&\dot{\pi}^{0}_{2}=\{\pi^{0}_{2},H_{T}\}=f^{-+}_{12}\approx0,\\
&\dot{\pi}^{2}_{2}=\{\pi^{2}_{2},H_{T}\}=-\omega^{-+}_{0,1}+\omega^{-2}_{1}\omega^{+2}_{0}-\omega^{-2}_{0}\omega^{+2}_{1}
+\lambda^{-+}_{1}-\Lambda\approx-\omega^{-+}_{0,1}-\omega^{-2}_{0}\omega^{+2}_{1}-\Lambda\approx 0,\\ 
&\dot{\pi}^{0}_{-+}=\{\pi^{0}_{-+},H_{T}\}=e^{2}_{2,1}-\omega^{-2}_{2}\approx0,\\
&\dot{\pi}^{1}_{-+}=\{\pi^{1}_{-+},H_{T}\}=e^{2}_{0,2}+\omega^{-2}_{2}e^{+}_{0}-\omega^{+2}_{2}\approx \dot e^2_2,\\ 
&\dot{\pi}^{2}_{-+}=\{\pi^{2}_{-+},H_{T}\}=-e^{2}_{0,1}+\omega^{-2}_{0}+\omega^{+2}_{1}-\omega^{-2}_{1}e^{+}_{0}\approx0,\\
&\dot{\pi}^{0}_{-2}=\{\pi^{0}_{-2},H_{T}\}=\omega^{-+}_{2}-\omega^{+2}_{1}e^{2}_{2}\approx0,\\ 
&\dot{\pi}^{1}_{-2}=\{\pi^{1}_{-2},H_{T}\}=-e^{+}_{0,2}+\omega^{+2}_{0}e^{2}_{2}-\omega^{-+}_{2}e^{+}_{0}-\omega^{+2}_{2}e^{2}_{0}\approx0,\\ 
&\dot{\pi}^{2}_{-2}=\{\pi^{2}_{-2},H_{T}\}=e^{+}_{0,1}-\omega^{-+}_{0}+\omega^{-+}_{1}e^{+}_{0}+\omega^{+2}_{1}e^{2}_{0}\approx0,\\ 
&\dot{\pi}^{0}_{+2}=\{\pi^{0}_{+2},H_{T}\}=\omega^{-2}_{1}e^{2}_{2}\approx0,\\ 
&\dot{\pi}^{1}_{+2}=\{\pi^{1}_{+2},H_{T}\}=\omega^{-+}_{2}-\omega^{-2}_{0}e^{2}_{2}+\omega^{-2}_{2}e^{2}_{0}\approx0,\\ 
&\dot{\pi}^{2}_{+2}=\{\pi^{2}_{+2},H_{T}\}=-\omega^{-2}_{1}e^{2}_{0}-\omega^{-+}_{1}\approx0.
\end{alignat}
The equation of motion of $e^{2}_{2}$, together with \eqref{32}, \eqref{34}, and \eqref{eq:psi5}--\eqref{eq:psi10}, constitutes the full set of torsion-free conditions.

\section{BTZ Spacetime}

 A simple well-known example is the BTZ spacetime,
whose metric can be written as
\begin{alignat}{1}
\od s^{2}=&-N^{2}\od v^{2}+2\od v\od r+r^{2}(\od\varphi+N^{\varphi}\od v)^{2}
\nonumber \\
=&[r^{2}(N^{\varphi})^{2}-N^{2}]\od v^{2}+2\od v\od r+2r^{2}N^{\varphi}\od v\od\varphi+r^{2}\od\varphi^{2},\\
N^{2}=&\frac{r^{2}}{\ell^{2}}-M+\frac{J^{2}}{4r^{2}},\qquad
N^{\varphi}=-\frac{J}{2r^{2}},
\end{alignat}
in Bondi-like coordinate, which is obviously a special form of the metric (\ref{eq:metric}).
The metric components are
\begin{alignat}{1}
g_{vv}=\frac{J^{2}}{4}-N^{2},\quad
g_{vr}=1,\quad g_{rr}=0,\quad
g_{v\varphi}=-\frac{J}{2},\quad g_{r\varphi}=0,\quad
g_{\varphi\varphi}=r^{2}.
\end{alignat}
The related coframe can be chosen as
\begin{alignat}{1}
&e^{-}=-\od v, \qquad
e^{+}=-\frac{N^{2}}{2}\od v+\od r,
\qquad
e^{2}=-\frac{J}{2r}\od v+r\od\varphi,
\end{alignat}
which is of course a special form of the coframe (\ref{eq:coframe-c2}) if one sets
\begin{alignat}{1}
e^{+}_{0}=&-\frac{N^{2}}{2},\qquad
e^{2}_{0}=-\frac{J}{2r},\qquad
e^{2}_{2}=r.
\end{alignat}
The coframe should also satisfy the torsion-free conditions
\begin{alignat}{1}
\od e^{I}+\omega^{IJ}\wedge e^{K}\eta_{JK}=0,
\end{alignat}
where $\omega^{IJ}$ are connections adapted with the above coframe.
Because of the torsion-free conditions, the connection components are dependent on the coframe components.
Therefore one can get the connections expressed by the coframe variables
\begin{alignat}{2}
\omega^{-+}_{0}=&-\frac{r}{\ell^{2}},\quad \omega^{-+}_{1}=0,\quad \omega^{-+}_{2}=\frac{J}{2r},
\quad \omega^{-2}_{0}=0,\quad \omega^{-2}_{1}=0,\quad \omega^{-2}_{2}=1,
\nonumber \\
\omega^{+2}_{0}=&N^{\varphi}N^{2},\quad \omega^{+2}_{1}=-N^{\varphi},\quad \omega^{+2}_{2}=-\frac{1}{2}N^{2},
\end{alignat}
Now one can use the above example to check whether the previous Hamiltonian analysis is correct or not.
One just needs to check whether the 12 secondary constraints are automatically satisfied in this special case, and the results are
\begin{alignat}{2}
\omega^{-2}_{1}=0, \quad \omega^{-+}_{1}=0, \quad \omega^{-2}_{2,1}=&0,
\quad \omega^{-+}_{2,1}+\omega^{-2}_{2}\omega^{+2}_{1}=(\frac{J}{2r})_{,1}-N^{\varphi}=0,
\quad e^{2}_{2,1}-\omega^{-2}_{2}=r_{,1}-1=0, 
\nonumber \\
e^{2}_{0,1}-\omega^{-2}_{0}-\omega^{+2}_{1}=&(-\frac{J}{2r})_{,1}+N^{\varphi}=0,
\quad \omega^{-+}_{2}-\omega^{+2}_{1}e^{2}_{2}=\frac{J}{2r}+N^{\varphi}r=0,
\nonumber \\
e^{+}_{0,2}-\omega^{+2}_{0}e^{2}_{2}+\omega^{-+}_{2}e^{+}_{0}+\omega^{+2}_{2}e^{2}_{0}
=&-N^{\varphi}N^{2}r-\frac{J}{2r}\frac{N^{2}}{2}+\frac{N^{2}}{2}\frac{J}{2r}=0,
\nonumber \\
e^{+}_{0,1}-\omega^{-+}_{0}+\omega^{+2}_{1}e^{2}_{0}=&(-\frac{1}{2}N^{2})_{,1}+\frac{r}{\ell^{2}}+N^{\varphi}\frac{J}{2r}=0,
\quad \omega^{-+}_{2}-\omega^{-2}_{0}e^{2}_{2}+\omega^{-2}_{2}e^{2}_{0}=\frac{J}{2r}-\frac{J}{2r}=0,
\nonumber \\
\omega^{-+}_{0,1}+\omega^{-2}_{0}\omega^{+2}_{1}+\frac{1}{\ell^{2}}=&(-\frac{r}{\ell^{2}})_{,1}+\frac{1}{\ell^{2}}=0,
\quad (e^{2}_{0,2}+\omega^{-2}_{2}e^{+}_{0}-\omega^{+2}_{2})_{,11}=(-\frac{1}{2}N^{2}+\frac{1}{2}N^{2})_{,11}=0,
\end{alignat}
which are all satisfied and prove the consistency of our analysis.


\section{Concluding Remarks}

The connection dynamics based on the decomposition, $\mathfrak{so}(1,d-1)= \mathfrak{so}(1,1)\oplus\mathfrak{so}(d-2)\oplus\mathfrak{t}^{-}(d-2)\oplus\mathfrak{t}^{+}(d-2)$, in a Bondi-like coordinate system is suggested.
The decomposition is valid for the Lorentz algebra $\mathfrak{so}(1,d-1)$ in an arbitrary
$d$-dimensional spacetime.  When an isolated horizon serves as the boundary of a spacetime,
the boundary SO$(1,1)$-BF theory can always be read out naturally from the variation of the action of gravity.  Besides, there is no signature problem in the remaining subalgebras, which might be easier to deal with than an $\mathfrak{so}(1,d-1)$ algebra in the quantization of the bulk system.

As a simple example, it is shown in the present paper that a self-consistent Hamiltonian formalism for the 3-dimensional $\frak{so}(1,1)\oplus\frak{t}^{-}\oplus\frak{t}^{+}$ connection dynamics can be set up in a Bondi-like coordinate system. In the system there are 12 independent primary constraints, 12 independent secondary constraints. All of them are second class constraints. There is no local physical degree of freedom in the system as expected.

In \cite{Dirac}, the consistency conditions are classified into three types.  The consistency conditions of the first type become identities on the constraint surface.  The second-type consistency-conditions provide the secondary constraints for the system.  From the consistency conditions of the third type, Lagrangian multipliers can be determined.  The
present paper shows that for a complicated constraint system, two solutions for the same Lagrangian multiplier may
be obtained from the consistency conditions. The two solutions may, in turn, give a new constraint, which is also a
secondary constraint. This situation was not discussed in the literature (see, for example, \cite{Dirac} and \cite{Li}).

It should be noted that there exists a local SO$(1,1)$ symmetry in an arbitrary dimensional gravitational theory.
It is the gauge symmetry in the direction of the generators of an isolated horizon and in the propagation direction of gravitational waves.
Due to the existence of SO$(1,1)$ symmetry, the local symmetry possess indefinite signature.  In the previous Hamiltonian analysis,
the $\frak{so}(1,1)$ sub-algebra does not appear in the decomposition of $\frak{so}(1,d-1)$ explicitly.
The decomposition of SO$(1,1)$ from SO$(1,d-1)$ may be useful to establish a method to analyze a
system in both Hamiltonian and Lagrangian formalisms.  The result of the paper
shows that the method is, at least, valid in 3-dimensional spacetime.

One may wonder why the three coordinate conditions are not considered in the Hamiltonian. In fact, a direct calculation shows that the multipliers
for the primary constraints of the coordinate conditions are all zero. In other words,
the addition of the coordinate conditions as new constraints will
not affect the constraint system. The only role is to make the analysis more complicated.

In the present analysis, the torsion-free conditions may be obtained automatically. Those without the term $e^I_{i,0}$ appear as secondary constraints,
those with the term $e^I_{i,0}$ appear in the Hamiltonian equations.  In an alternative way, one may treat
the torsion-free conditions as constraints at the beginning. However, it will not provide any new information.

It is remarkable that in the present formalism the metric on each slice ($v=const.$)
is always degenerate.  It is very different from the Hamiltonian analysis of gravitational theory in the literatures.
Although the Hamiltonian analysis in the Ashtekar formalism permits the degenerate geometry \cite{Rovelli,Ma-Liang-Kuang,Ma-Liang},
the non-degenerated geometry is mainly concerned.
However, what is degenerate in the formalism is just the induced geometry on each slice, while the 3-dimensional geometry is still non-degenerate.

It should be  finally remarked that if the parameters for gauge transformations are included in
the coframe fields at the beginning, the Hamiltonian analysis will give incorrect physical
degree of freedom.  This is caused by the absence of the transformation law of connection
in the analysis.  When the transformation law of connection is used, the correct physical degree of freedom will be recovered.

\section*{Acknowledgment}

We would like to thank Jing-Bo Wang for his helpful discussions.
This work is supported by National Natural Science Foundation of China under the grant
11275207 and 11690022.

\end{document}